\documentclass[aps,pra,twocolumn,superscriptaddress]{revtex4-1}

\usepackage[english]{babel}
\usepackage{amsmath}
\usepackage{amsthm}
\usepackage{amssymb}
\usepackage{mathrsfs}
\usepackage{booktabs}
\usepackage{color}
\usepackage{hyperref}
\usepackage[ruled,vlined]{algorithm2e}
\usepackage{cases}
\usepackage{dblfloatfix}
\usepackage{multirow}
\usepackage{xcolor}

\hypersetup{
  colorlinks   = true,  
  urlcolor     = blue,  
  linkcolor    = blue,  
  citecolor    = red    
}
\usepackage{mathtools}
\usepackage{natbib}
\usepackage{subfigure}
\usepackage{url}
\newcommand{\ket}[1]{ | #1 \rangle }
\newcommand{\bra}[1]{ \langle #1 | }

\newcommand{\z}{\mathbf{z}}

\mathtoolsset{showonlyrefs}

\DeclareMathOperator{\Tr}{Tr}

\theoremstyle{definition}

\theoremstyle{remark}

\theoremstyle{plain}
\newtheorem*{theorem*}{Theorem}

\usepackage{comment}
\usepackage{float}

\SetArgSty{textnormal}

\makeatletter

\newenvironment{widetext2}{%
  \par\ignorespaces
  \setbox\widetext@top\vbox{%
   \vskip15\p@
   \hb@xt@\hsize{%
    \leaders\hrule\hfil
    \vrule\@height6\p@
   }%
   \vskip6\p@
  }%
  \setbox\widetext@bot\hb@xt@\hsize{%
    \vrule\@depth6\p@
    \leaders\hrule\hfil
  }%
  \onecolumngrid
  \let\set@footnotewidth\set@footnotewidth@ii
}{%
  \par
  \twocolumngrid\global\@ignoretrue
  \@endpetrue
}%

\makeatother

\begin{document}

\title{Variational method for learning Quantum Channels via Stinespring Dilation on neutral atom systems}

\author{L.Y. \surname{Visser}}
\altaffiliation[Corresponding author: ]{l.y.visser@tue.nl}
\affiliation{Department of Mathematics and Computer Science, Eindhoven University of Technology, P.~O.~Box 513, 5600 MB Eindhoven, The Netherlands}

\author{R.J.P.T. \surname{de Keijzer}}
\affiliation{
Department of Applied Physics and Eindhoven Hendrik Casimir Institute, Eindhoven University of Technology, P. O. Box 513, 5600 MB Eindhoven, The Netherlands}

\author{O. \surname{Tse}}
\affiliation{Department of Mathematics and Computer Science, Eindhoven University of Technology, P.~O.~Box 513, 5600 MB Eindhoven, The Netherlands}

\author{S.J.J.M.F. \surname{Kokkelmans}}

\affiliation{
Department of Applied Physics and Eindhoven Hendrik Casimir Institute, Eindhoven University of Technology, P. O. Box 513, 5600 MB Eindhoven, The Netherlands}

\date{\today}

\begin{abstract}
Real-world quantum systems interact with their environments, leading to the \textit{irreversible} dynamics described by the Lindblad equation. Solutions to the Lindblad equation give rise to quantum channels $\Phi_t$ that characterize the evolution of density matrices as $\rho(t) = \Phi_t(\rho_0)$. In many quantum experiments, the observation windows are limited by experimental instability or technological constraints. Nevertheless, extending the evolution of the state beyond this window may be valuable for identifying sources of decoherence and dephasing or determining the steady state of the evolution. In this work, we propose a method to approximate {an arbitrary} target quantum channel by variationally constructing equivalent unitary operations on an extended system, leveraging the \textit{Stinespring dilation theorem}. We also present an experimentally feasible approach to extrapolate the quantum channel in discrete time steps beyond the period covered by the training data. Our approach takes advantage of the unique capability of neutral-atom quantum computers to spatially transport entangled qubits, an essential feature for implementing our method. The approach demonstrates significant predictive power for approximating non-trivial quantum channels.
\end{abstract}

\maketitle

\section{Introduction}
\label{sec:introduction}

In the noisy intermediate-scale quantum (NISQ) era \cite{Preskill_2018}, quantum computers suffer from noise and are limited in their computational power.
Nevertheless, these NISQ systems are useful for specific, well-designed tasks \cite{google}. Current predominant algorithms are variational quantum algorithms (VQAs) that aim to construct a unitary operation that minimizes a prescribed loss function. Certain VQAs have shown proof of concept for small dimensional problems with various hardware choices of qubits \cite{Kandala,trapion1,photonics1,majorana1,googlehartree}. Another application of NISQ-era quantum computers is quantum simulation. {Quantum computers have great potential for quantum simulations over classical computers, as they can capture quantum states without requiring resources that scale exponential with the problem size.} Here, the goal is to simulate the behavior of a closed quantum system evolving under the Schr\"{o}dinger equation, by mimicking the Hamiltonian of the target system. This results in a unitary evolution that fully describes the dynamics of the system.\\

On the other hand, many quantum systems of interest consist of a subsystem interacting with an environment, a so-called \textit{open quantum system}, e.g.\ photonic devices, fermionic lattices \cite{fermionopen}, ion collisions, quark-gluon plasmas \cite{Jong_2021} or even a NISQ quantum computer \cite{Wang2023}. This interaction leads to non-unitary evolutions described by the Lindblad equation \cite{lindblad1} (see~\eqref{eq:lindblad}), generally portraying dissipation in the form of dephasing and decoherence. The evolution operators associated with this equation are \textit{quantum channels}, the open system equivalent of unitary operators. {Deliberately simulating non-unitary behavior on a quantum system is non-trivial in most cases, as the native control is unitary and the dissipation that is present is generally uncontrollable noise from the environment. Nevertheless, studies are performed into using analogue quantum simulators to approximate the behavior of specific open quantum systems \cite{Kim_2022, Daley_2022, Georgescu_2014}. Instead, this paper proposes a digital approach using a quantum channel VQA that leverages the precise control and scalability of a quantum computer in combination with dilation theory. This allows for the approximation of the evolution of a wide class of open systems, which we call the \textit{target systems}}. \\

\begin{figure}[b]
    \centering    \includegraphics[scale=0.38]{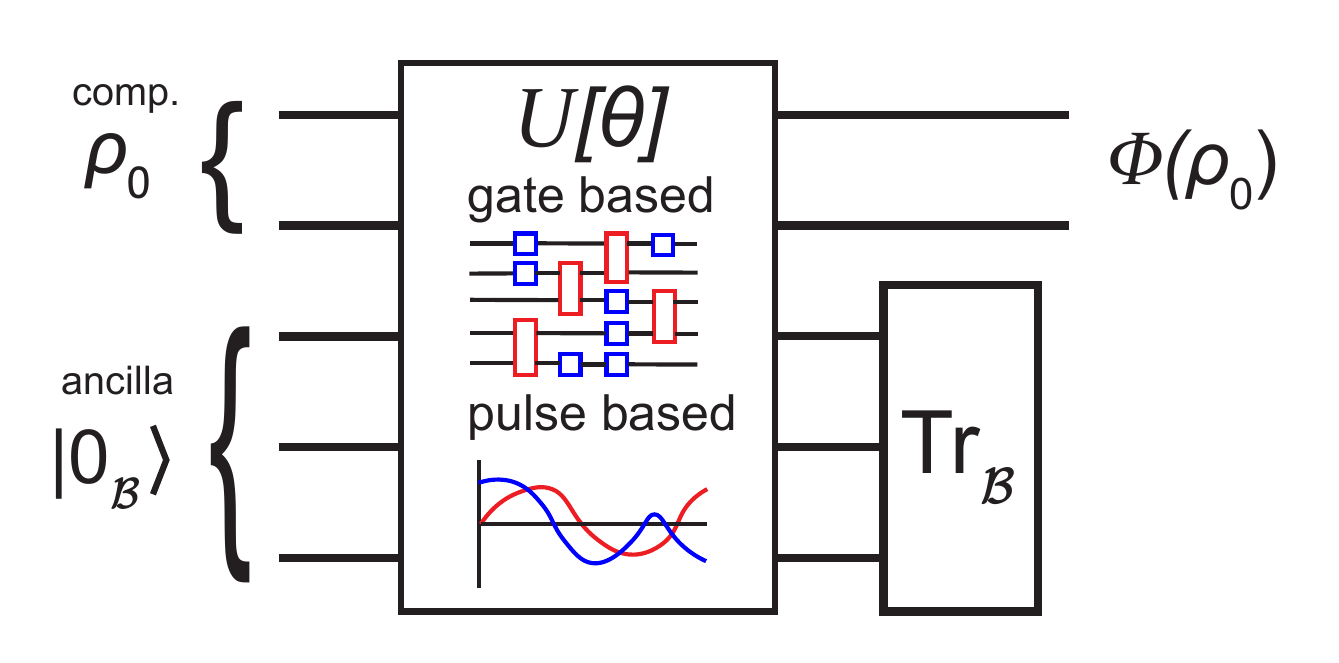}
    \caption{Schematic representation of a quantum channel $\rho_0\rightarrow\Phi(\rho_0)$ VQA using ancilla qubits $\mathcal{B}$. Both gate- and pulse-based methods can be used to approximate the Stinespring unitary $U[\theta]$.}
    \label{fig:quantumchannelvqa}
\end{figure}

\paragraph*{Approach.} Our method {uses a quantum computer to learn the non-unitary evolution at discrete time steps of a target system using measurement data, based on dilation theory \cite{shalit2020dilation}.} The method first considers a number of computational qubits on the quantum computer, given the dimensionality of the Hilbert space in which the target quantum channel (tqc) is described. To capture the behavior of the environment, this Hilbert space is extended by introducing a number of ancilla qubits based on the Stinespring dilation theorem (see Sec.~\ref{sec:Stinespring}) \cite{stinespring7}. The method then learns a unitary operator on this extended system (computational + ancilla qubits)--- called the \textit{Stinespring unitary}--- using a VQA based on input measurement data on the computational qubits. The ancilla qubits (environment) are then traced out of the Stinespring unitary to provide an approximation of the target quantum channel; see Fig.~\ref{fig:quantumchannelvqa}.\\

By recursively storing away the old ancilla qubits and applying the Stinespring unitary on the computational qubits and a new set of ancillas, the application of the quantum channel can be repeated, resulting in approximations of the target quantum channel at discrete time steps, {similar to the approach of \textit{collision models} \cite{ciccarello2022}. These models are a specific type of system-bath model used in the investigation of open quantum systems}. Because of its ability to coherently move qubits around \cite{Bluvstein_2022, Lukin2023}, a neutral atom quantum computing system is especially well-suited for initializing a new set of ancilla qubits. {In this work, we further explore the implementation of the Stinespring dilation method on neutral atom quantum computers.} \\

\paragraph*{Relation to previous work.} {
Several other works approximate quantum channels by applying dilation theorems on \textit{known} quantum channels and implementing the resulting unitary dilation on quantum computers \cite{gaikwad, Hu_2018, Hu_2020, Marsden_2021, Hu_2022, Jong_2021, stinespring7,Wang2023, GarciaPerez2020,Schlimgen_2022}, thus requiring prior knowledge of the quantum channel operators. Specifically, Refs.~\cite{Jong_2021, Hu_2018} are based on Stinespring's dilation theorem \cite{stinespring_1954_original} while Refs.~\cite{gaikwad, Wang2023,stinespring7, Hu_2022, Hu_2020,Marsden_2021} use the Sz.~Nagy dilation \cite{sznagy} or unitary decomposition methods, applied either to the corresponding Kraus or Lindblad operators (see~\eqref{eq:lindblad}). Explicit constructions of unitary dilations have been obtained for specific use cases \cite{Jong_2021, Hu_2022, fermionopen}. In contrast to this, our method invokes quantum state tomography \cite{giacomo2023, Isaac1997, Ahmed2023, Mohseni_2008} to determine the properties of a quantum channel based on measurements.
}\\ 

{
Closely related to these dilation methods are other ancilla qubit methods for quantum channel simulation, including quantum Zeno dynamics \cite{quantumzeno}, which is a variation of our method where the ancilla qubits get measured every iteration instead of stored; unitary decomposition methods \cite{unitarydecomposition}, where the Kraus operators are decomposed into linear combinations of unitary operators; and imaginary time evolution \cite{fermionopen,natureimaginary}, where the non-unitary dynamics of imaginary time evolutions are simulated by recursively updating parameters of an ansatz.}\\

{
In particular, Sz.~Nagy and unitary decomposition methods give a constructive manner of creating the quantum channel, either involving a large ancilla space or practically difficult to implement dilation, as in Refs.~\cite{Marsden_2021, Hu_2020}, or very specific use cases, as in Refs.~\cite{Jong_2021, Hu_2022, fermionopen}. The Stinespring dilation theorem merely implies the existence of a unitary dilation, which can consequently be learned variationally using experimental data. To learn a practically implementable unitary dilation, our method invokes quantum state tomography \cite{giacomo2023, Isaac1997, Ahmed2023, Mohseni_2008} to determine the properties of a quantum channel based on measurements.}\\

{Extrapolating quantum channel behavior in line with the theory of collision models has been studied in \cite{ Cuevas2019, Cech2023, Cilluffo2021}. In contrast to our method, \cite{Cech2023, Cilluffo2021} use collision models to approximate \textit{known} quantum channels using measurements of the ancilla qubits, creating a stochastic quantum trajectory. This avoids the need to maintain coherence, thus limiting the method. The recent work of Ref.~\cite{cemin2024} most closely resembles our approach by variationally learning dilation operators to characterize the noise inherent in the quantum computing system. Our method extends beyond this by generalizing to the simulation of arbitrary quantum systems.}\\

{We believe our work is the first to combine the Stinespring dilation theorem with quantum state tomography for learning quantum channels and uses collision models to extrapolate said channels. We stress that the resulting method is agnostic to the underlying target quantum channel and can be practically implemented on a pulse-based neutral atom system.}\\

The layout of this paper is as follows. Sec.~\ref{sec:Stinespring} describes quantum channels and the Stinespring dilation theorem. Sec.~\ref{sec:vqas} details the quantum channel VQA developed here. Sec.~\ref{sec:neutralatoms} describes how the algorithm is readily tailored toward execution on a NISQ neutral atom quantum computing system. In Sec.~\ref{sec:results}, we show the initial results of our quantum channel approximation method and compare gate- and pulse-based methods. 

\section{Quantum Channels \& Stinespring Dilation}
\label{sec:Stinespring}

\begin{figure}[t]
    \centering
    \includegraphics[scale=0.5]{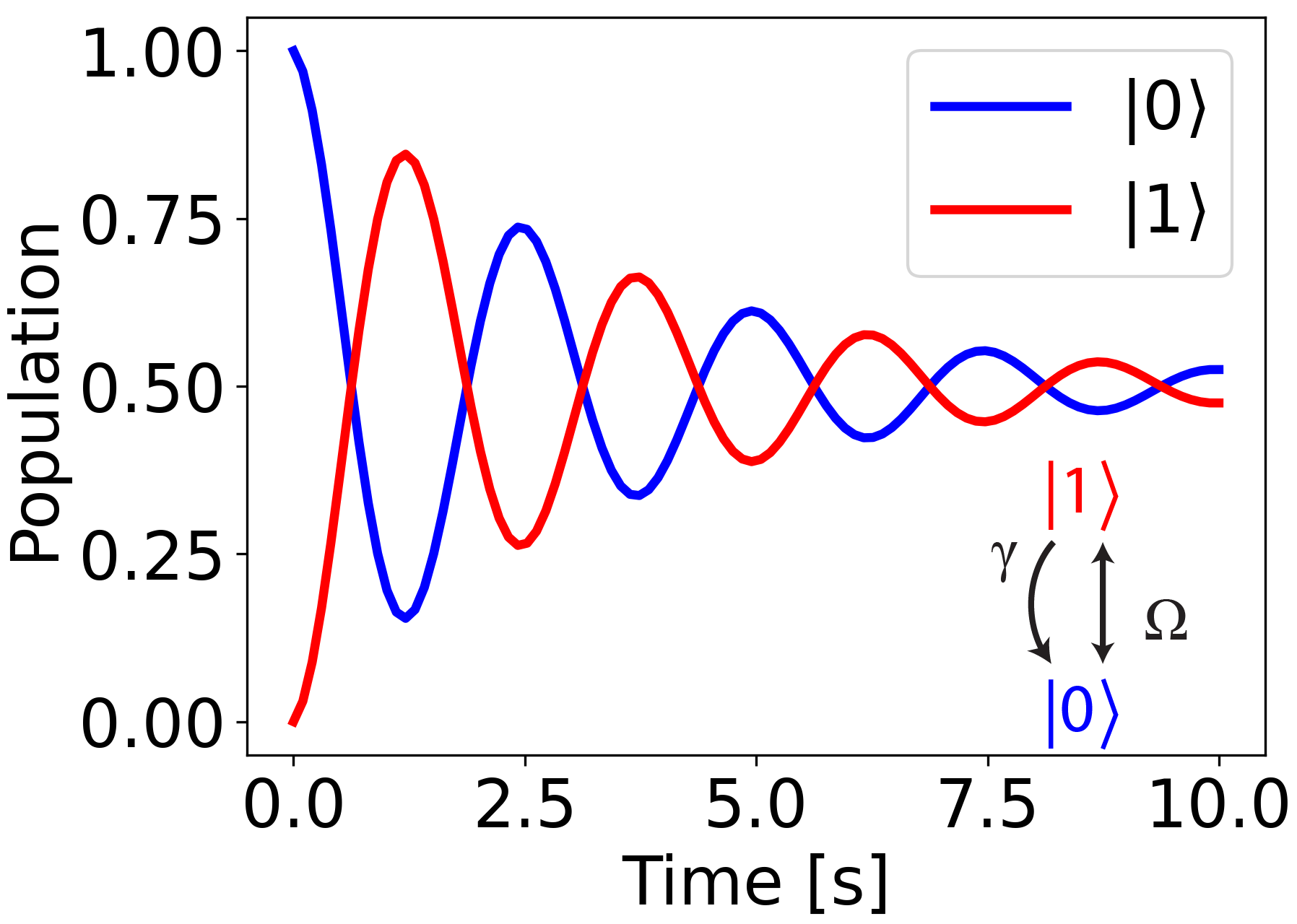}
    \caption{Example of quantum channel behavior showing populations for single qubit Rabi oscillation with $\Omega=0.4 \pi$ Hz, $\Gamma=|0\rangle\langle 1|$ and $\gamma=0.15$ Hz decay.}
    \label{fig:singlequbitdecayarb}
\end{figure}

Most quantum systems in nature interact with their environment. This results in an open quantum system that undergoes decoherence and dephasing \cite{decoherence1} (see Fig.~\ref{fig:singlequbitdecayarb}), which prevents them from being represented by pure states $|\psi\rangle$, and requires the use of density matrices $\rho$, henceforth called states. The evolution of states in such an open quantum system can be described by the Lindblad equation \cite{lindblad1,lindblad2}
\begin{equation}
\label{eq:lindblad}
    \partial_t \rho = -i \left[H_\mathcal{A}, \rho \right] + \sum_k \gamma_k \Gamma_k \rho \Gamma_k^\dag - \frac{1}{2} \gamma_k \left\{\Gamma_k^\dag \Gamma_k, \rho \right\},
\end{equation}
with a (time-independent) system Hamiltonian $H_\mathcal{A}$ acting on the system of interest, jump operators $\Gamma_k$ with corresponding decay rates $\gamma_k$ that characterize the interactions with the environment, and $[\cdot,\cdot]$ and $\{\cdot,\cdot\}$ being the commutator and anti-commutator respectively. Solutions to the Lindblad equation are described by a quantum channel $\Phi_t$ for each time $t$ acting on states $\rho_0$ as
\begin{equation}
    \rho(t) = \Phi_t(\rho_0),
\end{equation}
with $\Phi_t$ being linear, trace preserving, and completely positive, i.e., $I_n \otimes \Phi_t$ is positive for every $n\in \mathbb{N}$ \cite{holevo}.

The qubits of an ideal quantum computer form a closed system and evolve under unitary transformations. As a result, it is impossible to mimic a non-unitary quantum channel $\Phi$ acting on a $2^m$ dimensional space by using a quantum computer with only $m$ qubits. In dilation theory, operators on a Hilbert space $\mathcal{A}$ are extended as projections of operators that act on a larger Hilbert space $\mathcal{A}\otimes \mathcal{B}$. The new operator is then called a dilation of $\Phi$ and can have more favorable properties than the original operator \cite{shalit2020dilation, Evans_1976, stinespring_1954_original}.  For this purpose, we employ the Stinespring dilation theorem \cite{stinespring2} applied to quantum channels, which we recall here.

\begin{theorem*}[Stinespring dilation theorem]\label{thm:stinespring dilation}
    Let $\mathcal{A}$ be a finite dimensional Hilbert space and $\Phi: \mathcal{A} \to \mathcal{A}$ be a quantum channel. Then there exists a Hilbert space $\mathcal{B}$ with $\dim(\mathcal{B}) \leq \dim(\mathcal{A})^2$ and a unitary $U: \mathcal{A}\otimes \mathcal{B} \to \mathcal{A} \otimes \mathcal{B}$ such that
    \begin{equation}
        \Phi(\rho) = \Tr_\mathcal{B}\bigl[U \cdot\rho \otimes \ket{0}_\mathcal{B} \bra{0}_\mathcal{B} \cdot U^\dagger\bigr].
    \end{equation}
    where $|0\rangle_\mathcal{B}$ is the zero state on $\mathcal{B}$.
\end{theorem*}

The Hilbert space of a quantum computer with $m$ qubits has a dimensionality of $2^m$. As a result, at most $3m$ qubits are needed to simulate a quantum channel on a $2^m$ dimensional system. Moreover, it has been shown that $\dim(\mathcal{B}) = k$, where $k$ is the number of jump operators present in the Lindblad equation of~\eqref{eq:lindblad} \cite{Hu_2020}. Generally, we call the qubits that live in the space $\mathcal{A}$ computational qubits and the qubits that live in the space $\mathcal{B}$ ancillary qubits. Note that for $k$ jump operators, a minimum of $\log_2(k)$ additional ancillas are required.\\

Knowledge of the unitary dilation of a quantum channel at the initial sample time $\Delta t$ opens up the possibility of extrapolation if all underlying processes $H_\mathcal{A}$ and $\gamma_k$ are time independent. Indeed, if the quantum channel is given by $\rho(\Delta t) = \Phi_{\Delta t}(\rho_0)$, then the semigroup property holds, i.e. for every $n\in\mathbb{N}$
    \begin{equation}
 \label{eq:multitimestep}
        \rho(n \Delta t) = \underbrace{\Phi_{\Delta t} ( \Phi_{\Delta t} ( \cdots \Phi_{\Delta t}(\rho_0) \cdots ))}_{n  \text{ repeated applications}}=\Phi_{n\Delta t}(\rho_0).
    \end{equation}

Moreover, the Stinespring dilation theorem provides the existence of a unitary transformation $U$ satisfying
\begin{equation}
\label{eq:dilationt}
    \rho(\Delta t) = \Phi_{\Delta t}(\rho_0) = \Tr_{\mathcal{B}}\bigl[U \cdot \rho_0 \otimes \ket{0}_\mathcal{B} \bra{0}_\mathcal{B} \cdot U^\dagger\bigr],
\end{equation}
and consequently,
\begin{equation}
    \rho((n+1) \Delta t) = \Tr_\mathcal{B}\bigl[U \cdot \rho(n \Delta t) \otimes \ket{0}_\mathcal{B} \bra{0}_\mathcal{B} \cdot U^\dagger\bigr].
\end{equation}
Thus, if an approximation for the behavior of a quantum channel is known at a fixed time $t=\Delta t$, then the approximation can be reapplied $n$ times to obtain predictions of the state at time $t=n \Delta t$. \\

When reapplying the Stinespring unitary, one cannot use the same set of ancillas, as the old ancillas have become entangled with the computational qubits by application of the unitary $U$. This means that any disturbance of the used ancillas also disturbs the state of the computational qubits. This calls for a new set of ancillas to be brought in from a reservoir, while the old set of ancillas has to be coherently stored. In Sec.~\ref{sec:neutralatoms}, we detail how a neutral atom quantum computing system's ability to coherently move around entangled qubits is well suited for this purpose \cite{Bluvstein_2022}.

\section{Variational quantum algorithms}
\label{sec:vqas}

\subsection{Input data and loss function}
The goal of our method is to approximate the quantum channel $\Phi = \Phi_{\Delta t}$ by a parametrized quantum channel $\widehat{\Phi}[\theta]$ taking the form
\begin{equation}
\label{eq:dilationt2}
 \widehat{\Phi}[\theta](\rho_0) = \Tr_\mathcal{B}\bigl[U[\theta] \cdot \rho_0 \otimes \ket{0}_\mathcal{B} \bra{0}_\mathcal{B} \cdot U^\dagger[\theta]\bigr],
\end{equation}
where the Stinespring unitary $U[\theta]$ is constructed on a quantum computer by training for the parameters $\theta$ given a loss function and input data.  {In problem-specific quantum channel simulators (as mentioned in Sec.~\ref{sec:introduction}), knowledge about the intricacies of the quantum channels is assumed, e.g.\ the exact form of the Kraus operators. This allows for a theoretical model for dilation $U$, which can then be implemented using the fine-tuned Hamiltonian evolution of a quantum simulator \cite{Kim_2022,Georgescu_2014}. Our method differentiates itself from this by learning the dilation $U$ for a large set of quantum channels, purely from measurement data.}  A set of $L\in\mathbb{N}$ pairs of initial states and their evolution under the target quantum channel $\{\rho_{l,0}, \rho_{l,1}\coloneqq \Phi(\rho_{l,0})\}_{l=1}^L$ would be very desirable input data. However, in many experiments $\rho_{l,1}$ will not be fully known. Instead, information on $\rho_{l,1}$ is only known through measurements against an observable $O_l$, giving $\Tr[O_l \rho_{l,1}]$. In this case, the input data is given by $\{\rho_{l,0},\Tr[O_l\rho_{l,1}]\}_{l=1}^L$. Note that $\rho_{l,0}$ can be taken identically for different observables $O_l$. Based on this, we set the loss function as
\begin{equation}\label{eq:lossfunc}
   J(U) \coloneqq \sum_{l=1}^L \Bigl(\Tr_A[O_l \widehat{\rho}_{l,1}] - \Tr_A[O_l \rho_{l,1}]\Bigr)^2,
\end{equation}
where $\widehat{\rho}_{l,1}:=\widehat{\Phi}(\rho_{l,0})$.\\

The Pauli strings are a good choice for the observables, as the set of Pauli strings forms a basis of all Hermitian operators. Thus, if $\{O_l\}_{l=1}^L$ contains all Pauli strings for every unique $\rho_{l,1}$, a zero loss in~\eqref{eq:lossfunc} corresponds to the identity $\tilde{\rho}_{l,1} = \rho_{l,1}$ for all $l$. However, if not all Pauli strings are taken into account, the states do not necessarily have to match for the missing Pauli strings. \\

Appendix~\ref{app:multistep} provides details on how to extend our methods for input data on $n$ repeated applications of the quantum channel $\Phi$, i.e.\ $\{\rho_{l,0},\Tr[O_l\rho_{l,1}],\Tr[O_l\rho_{l,2}],...\}_{l=1}^L$ with $\rho_{l,n} = \Phi(\rho_{l,n-1})$. {As the original quantum channel $\Phi$ corresponded to a time step of $\Delta t$, applying $\Phi$ up to $n$ times corresponds to a time evolution of at most $n \Delta t \eqqcolon t_{\text{train}}$. If the density matrix $\rho_{l,n}$ can be fully determined from the measurement of observables $O_l$, the repeated application only gives more pairs of input-output to train the quantum channel $\widehat{\Phi}$. However, if fewer observables are measured, the error in repeated applications can be used to infer information about the channel that is not captured directly by the observables.} If the steady state $\rho_\infty=\Phi_t(\rho_\infty)$, $t>0$ is known, this can also be included as input data. In the rest of this section, we describe the optimization of $\theta$ based on a gate- and pulse-based quantum state evolution method. {The unitary dynamics described by $U[\theta]$ has an associated timescale determined by the underlying hardware of the quantum computer. We note that the dynamics of the target quantum channel arise from different dynamics and are independent of this. To make a clear distinction between these two timescales, we will use $\tau$ exclusively for the time-related quantities on the quantum computer and $t$ for the target quantum channel.}\\

\subsection{Gate-based optimization}
One way to train for the Stinespring unitary $U$ is using a parametrized gate sequence. A commonly used template for such a gate sequence is the hardware-efficient ansatz \cite{Kandala,ansatzes}. This sequence alternates between blocks of parametrized single qubit gates $U_{q,j}$ executed in time $\tau_g$, with $q$ indicating the qubit and $j$ the block, and a fixed entangling gate $U_{ent}$. An easy way to implement such an entangling gate is to let the system evolve for a time $\tau_V$ under its drift Hamiltonian $H_V$ (the passive evolution of the system, see App.~\ref{app:rydberg}) such that $U_{ent}=\exp(-i \tau_V H_V)$. The Stinespring unitary $U[\theta]$ as in~\eqref{eq:dilationt2} then takes the form
\begin{equation}
\label{eq:equationstatepreperation}
\begin{aligned}
    U[\theta]=&\underbrace{\left[U_{ent}\prod^m_{q=1} U_{q,d}(\theta) \right]\cdots\left[U_{ent}\prod^m_{q=1} U_{q,1}(\theta) \right]}_{d \text{ times}},
\end{aligned}
\end{equation}
where the depth $d$ of a state preparation is defined as the number of blocks in the gate sequence. The total execution time is, therefore, $\tau_{f,\text{gate}}=d(\tau_g+\tau_V)$ with $d$ the depth of the hardware-efficient ansatz. This method is especially relevant in NISQ machines, where single qubit gates $U_{q,d}$ can be implemented with high fidelity \cite{rydberg1}. If some form of control is present in entanglement operations, there is freedom to choose $U_{ent}$, which is influential on the performance of the algorithm \cite{entanglinggateinfluence}.\\ 

The parameters $\theta$ are optimized by gradient descent using finite differences. More sophisticated methods for gate-based optimization using analytic gradients exist \cite{schuld}, which may be included in future work. The number of parameters for a hardware-efficient Ansatz is $3\cdot d\cdot m$. Thus, the gate-based algorithm requires $\#\text{QE}=3\cdot d\cdot m$ quantum states evaluations to find the gradient using finite differences. {The number of quantum evaluations will be important when comparing gate-based and pulse-based methods in Sec.~\ref{sec:pulsebasedgatebased}. 
We heuristically find that a stochastic gradient descent \cite{stochasticgradient}, where a random batch of parameters is updated in each iteration, reduces the number of quantum evaluations required to converge.}

\subsection{Pulse-based optimization}\label{sec:pulsebasedopt}
Alternatively, Stinespring-based unitaries can be approximated via pulse-based optimization approaches \cite{QSLpulse,qocvqe1,pulsesandbackagain,deKeijzer2023pulsebased}. This approach takes an analog approach to state preparation. Pulse-based approaches have the advantages of faster state preparation and higher expressibility of the Hilbert space, both of which are important factors for the mitigation of decoherence in the NISQ era \cite{deKeijzer2023pulsebased}. \\

The goal of pulse-based optimization is to solve the minimization problem
\begin{equation}
    \min_{\z = (z_1,\ldots,z_R)} J(U[\z]) + \frac{\lambda}{2}\sum_{r=1}^R \int_0^{\tau_f} |z_r(\tau)|^2  d\tau,
\end{equation}
for complex-valued square-integrable control pulses $z_r\in L^2([0,\tau_f],\mathbb{C})$, and where the controlled unitary $U[\z]$ satisfies the Schr\"{o}dinger equation as
\begin{equation}
\label{eq:hamiltoniaoc}
    i\partial_\tau U = \big(H_V+H_c[\z]\big) U, \quad U(0)=I.
\end{equation}
Here, $\tau_f>0$ is the pulse end time, and $\lambda>0$ is a regularization parameter penalizing the total pulse energy. The Hamiltonian is split up into an uncontrolled part $H_V$ (the drift Hamiltonian) and a controlled part $H_c[\z]$ (the control Hamiltonian), see App.~\ref{app:rydberg}, which takes the form
\begin{equation}
\label{eq:controlhamiltonian4}
    H_c[\z]=\sum_{r=1}^R z_r Q_r +\overline{z_r} Q_r^\dagger,
\end{equation}
for a set of control operators $Q_r$, $r = 1, \ldots, R$.\\

These minimization problems are solved using the optimal control framework as in Ref.~\cite{deKeijzer2023pulsebased}, leading to the following KKT optimality conditions
\begin{equation}
\begin{aligned}
\label{eq:kkt}
     i \partial_\tau U - \bigl(H_V + H_c[\z] \bigr) U(\z) &= 0, \\
     \lambda z_r + \Tr\bigl[{Q_r^\dag\left(P U(\z)^\dag + U(\z) P^\dag \right)}\bigr] &= 0, \\
    i\partial_\tau P -  \bigl(H_V^\dag + H_c[\z]^\dag \bigr) P &= 0,
\end{aligned}
\end{equation}
where $P(\tau)$, $\tau\in[0,\tau_f]$ is the \textit{adjoint process} satisfying the terminal condition
\begin{equation}
\label{eq:frechet J1}
\begin{aligned}
    P(\tau_f) =& -4  i \sum_l \Tr_\mathcal{A}\bigl[O_l ( \tilde{\rho}_{l,1} - \rho_{l,1})\bigr]\\ &\times \Bigl(O_l \otimes I_\mathcal{B} \cdot U(\tau_f) \cdot \rho_{l,0}\otimes \ket{0}\bra{0}_\mathcal{B}\Bigr).
\end{aligned}
\end{equation}
It is easily shown that
\begin{equation}
    P(\tau) = U(\tau) U(\tau_f)^\dag P(\tau_f), \quad \tau\in[0,\tau_f],
\end{equation}
satisfies the requirements on $P$. The trace term in the second equation of \eqref{eq:kkt} may be expressed as
\begin{equation*}
\begin{aligned}
    \eta_r(\tau) & \coloneqq \Tr_{\mathcal{A}\otimes \mathcal{B}}\Bigl[Q_r^\dagger\Bigl(P(\tau)U^\dagger(\tau) + U(\tau)P^\dagger(\tau)\Bigr)\Bigr]\\
    &= \sum_{l}\sum_{k=1}^K 4i\Tr_{\mathcal{A}}[O_l(\widehat{\rho}_{l,1}-\rho_{l,1})] \\&\times\Tr_{\mathcal{A}\otimes \mathcal{B}}\left[ \widehat{\rho}(\tau)\Big[V_{k,r}^\dagger,\, \Gamma^\dagger(\tau_f,\tau) (O_l\otimes I_\mathcal{B}) \Gamma(\tau_f,\tau)\Bigr]\right],
    \label{eq:470}
\end{aligned}
\end{equation*}
where $\widehat{\rho}(\tau):=U(\tau)(\rho_{l,0}\otimes |0\rangle\langle 0|_\mathcal{B})U(\tau)^\dagger$ is the evolved state of the extended system up to time $\tau$, $\Gamma(\tau_f,\tau):=U(\tau_f)U^\dagger(\tau)$ describes the evolution by the pulses from $\tau$ to $\tau_f$, $[\cdot,\cdot]$ is the usual commutator, and $V_{k,r}$ are unitaries decomposing {the Hermitian control operator} $Q_r$ as
\begin{equation*}
Q_r=\sum_{k=1}^K V_{k,r},\quad r=1,...,R,
\end{equation*}
where $K\in\mathbb{N}$ is the necessary number of unitaries. For practical purposes, where the control terms work on only 1 (or occasionally 2) qubits, $K=O(1)$. These terms can be efficiently determined on a quantum computer, by first applying the pulse until time $\tau$, performing a single gate operation $V_{k,r}$, then applying the rest of the pulse up to $\tau_f$ and finally measuring the expectation of $O_l$ (cf.~\cite{deKeijzer2023pulsebased}). The pulses are iteratively updated as 
\begin{equation}
    z_{r,k+1}(\tau)=z_{r,k}(\tau)-\alpha_k(\lambda z_{r,k}(\tau)+\eta_r(\tau)),
\end{equation}
where $\alpha_k$ is a step size, in this work determined using the Armijo condition \cite{armijo} and $z_{r,0}(\tau)$ are constant zero value pulses.\\

The pulses are discretized as equidistant piecewise constant functions with $N\in\mathbb{N}$ steps, resulting in $\#\text{QE}=N\cdot K\cdot R$ quantum evaluations per iteration. We refer to \cite{deKeijzer2023pulsebased} for further details on the pulse-based method.

\subsection{Barren plateaus \& Hamiltonian splitting}\label{sec:BarrenPlateauHamiltonianSplitting}
Similar to most VQAs, the training of the Stinespring unitary for a large computational system suffers from barren plateaus \cite{barrenplateau1,barrenplateau2, Cunningham24}, i.e., large parts of parameter space for both gate- and pulse ansatzes have gradients that are exponentially small in system size \cite{deKeijzer2023pulsebased}. As mentioned in Sec.~\ref{sec:Stinespring}, the maximal number of necessary ancillas is twice the number of system qubits. Thus, as the total numbers of qubits is still linear in the system size, quantum channel learning has the same exponential scaling inherent to barren plateaus as unitary learning. \\

Due to the promise of VQA's, the barren plateau phenomenon has been widely explored to better understand the associated problems and to develop methods to circumvent those \cite{Cunningham24, barrenplateau2, barrenplateauansatz, Cerezo21, localminima, Pesah21, Wang20}. In Ref.~\cite{Cerezo21}, it was found that shallow circuits with local cost functions do not suffer from the barren plateau phenomenon. However, shallow circuits might lack the required expressibility and get trapped in local minima \cite{localminima}. Barren plateaus will be absent for specific cost functions, as is shown for state preparation using the quantum Wasserstein distance \cite{Kiani_2022}. Furthermore, it has been shown that optimizing over the Fourier coefficients of a pulse-based ansatz can mitigate barren plateaus in ground state problems \cite{Broers24}. Finally, there exists initialization schemes and parameter-constrained ansatzes for VQAs that are free from barren plateaus \cite{barrenplateauansatz, Pesah21}. \\

\begin{figure}
    \centering
    \includegraphics[width = 0.8\linewidth]{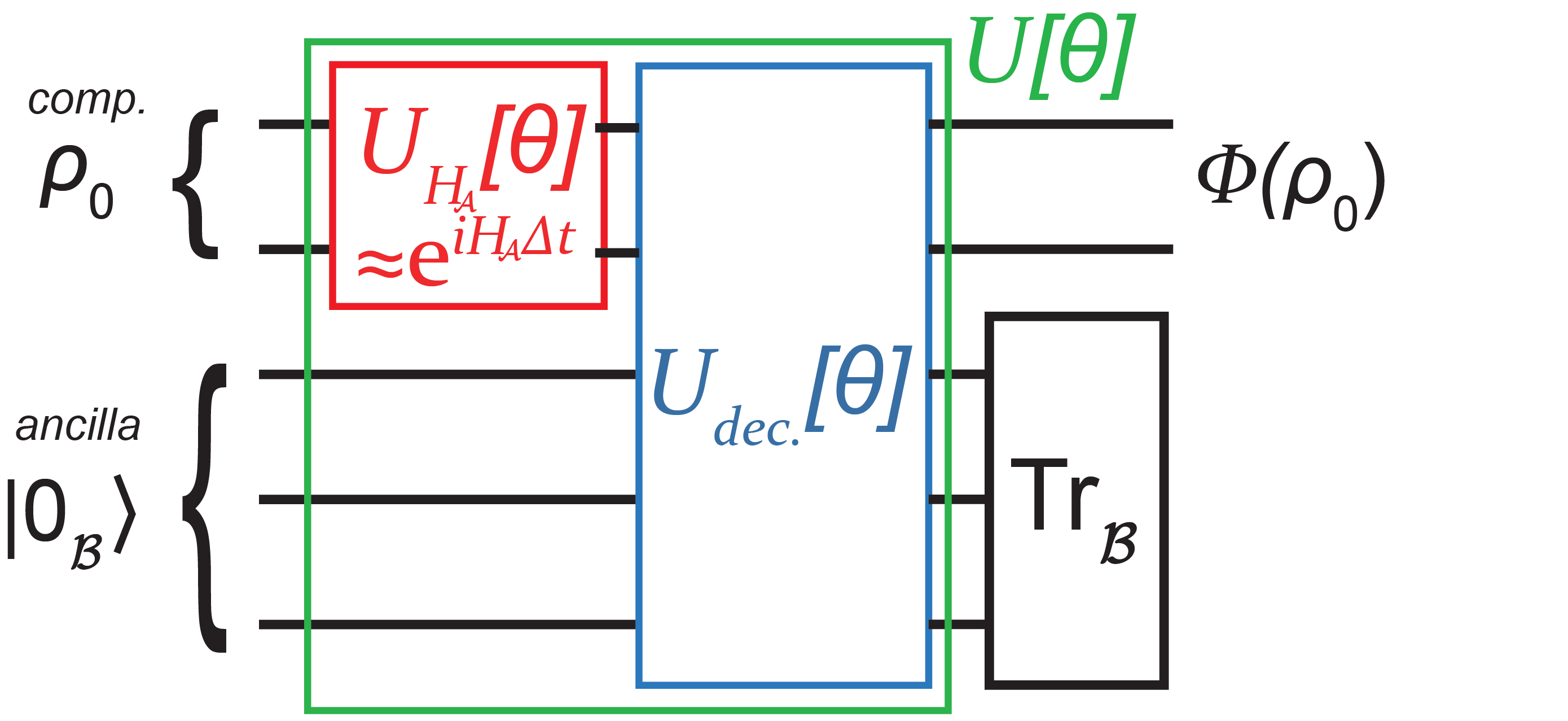}
    \caption{Decomposing the unitary $U$ into a Hamiltonian part $U_H$ and a decoherence part $U_{\text{dec.}}$ has the potential to lead to faster optimization of the parameters $\theta$, which is beneficial in terms of scalability and trainability.}
    \label{fig:hamiltoniansplitting}
\end{figure}

For the quantum channel approximation, a possibly interesting ansatz for barren plateau mitigation is to split the Stinespring unitary into a part that is unitary on the computational qubits and a dilation part, as in Ref.~\cite{Jong_2021}. The unitary part takes the role of the Hamiltonian evolution of \eqref{eq:lindblad}, while the dilation models the decoherence. By separating these two contributions and thus training two unitaries (as in Fig.~\ref{fig:hamiltoniansplitting}), the model could potentially become less complex and better suited for cases where the unitary evolution and decoherence act on different timescales. Moreover, using a reasonable ansatz for the Hamiltonian evolution (which is often better understood) can be especially beneficial for the scalability and trainability of the method. The reason for this is that the unitary part encoding the Hamiltonian evolution is of a much lower dimension, and thus training within this space suffers substantially less from the many local minima in shallow-circuit VQAs \cite{localminima}. Furthermore, for the dilation part, it is possible to set a good ansatz for a single decay channel, as seen in App.~\ref{app:1qubitdecay}. In a general setting with minimal knowledge of the underlying system, this split ansatz will not be good enough to circumvent the barren plateau. Solving the barren plateau problem for those type of settings is outside the scope of this work.

\section{Execution on neutral atom systems}
\label{sec:neutralatoms}
Neutral atom quantum computing architectures have emerged as a promising candidate for NISQ computing, reaching single-qubit gate fidelities of 0.9996 \cite{fidelitythompson} and two-qubit gate fidelities of 0.995 \cite{fidelitylukin}. In this system, the qubits are individual atoms trapped in laser optical tweezers. These tweezer sites can be moved and rearranged using AOM techniques \cite{madjarov}, resulting in adaptable qubit geometries. Entanglement in neutral atom systems is supplied by excitations to high-lying Rydberg states, which interact using Van der Waals interactions \cite{review2}. The electronic states of the atoms encode the qubit states. One possibility is the implementation of ground-ground qubits, in which two (meta-)stable states are chosen as the $\{|0\rangle,|1\rangle\}$ qubit manifold and a Rydberg state $|r\rangle$ is seen as an auxiliary state used for entanglement. On the other hand, the (often less stable \cite{dekeijzer2023recapture}) Rydberg state can assume the role of the $|1\rangle$ qubit state, which results in ground-Rydberg qubits \cite{review1,review3}. \\

\begin{widetext2}
    \begin{minipage}[b]{\linewidth}
    \begin{figure}[H]
            \centering
            \includegraphics[scale=0.307]{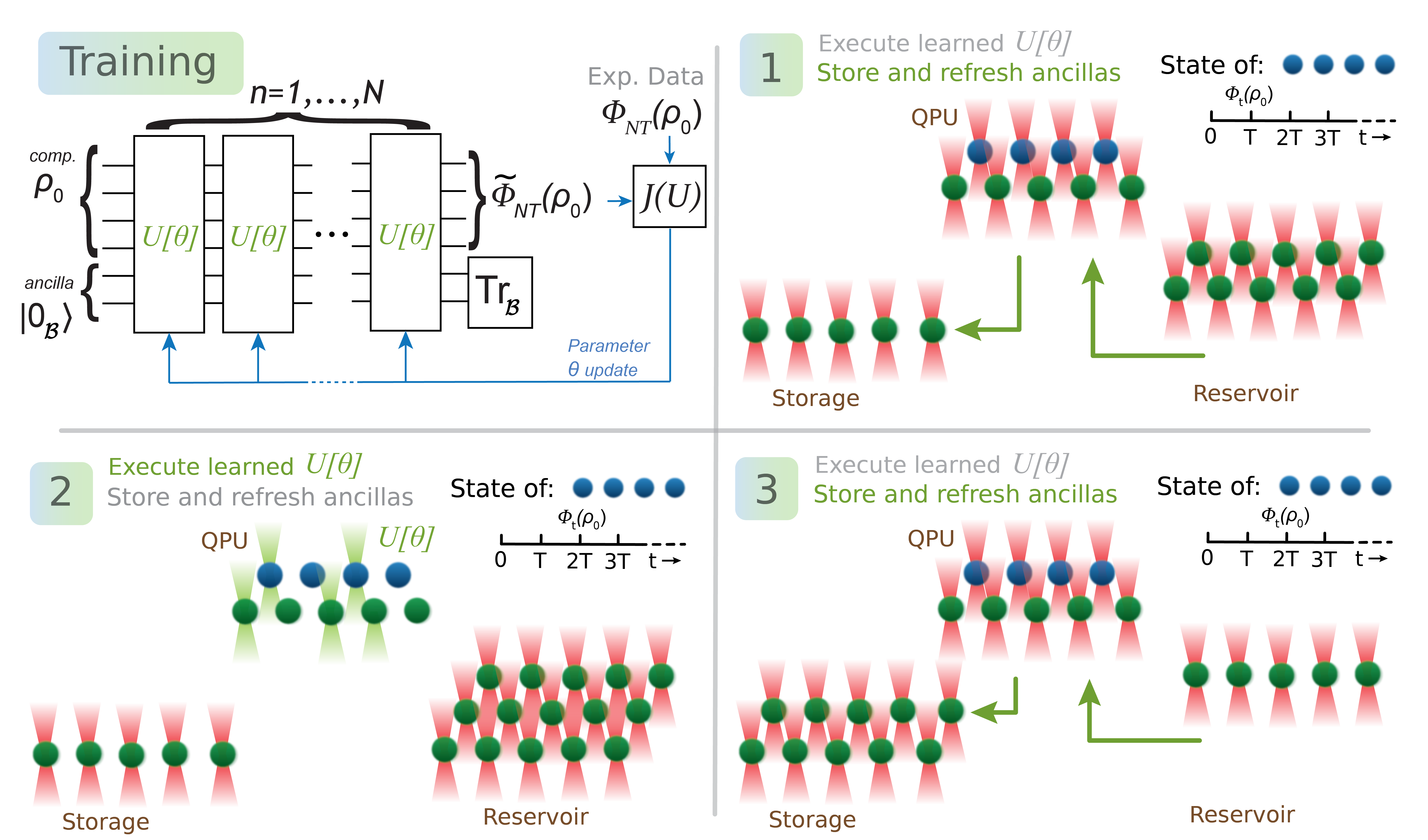}
            \caption{A neutral atom implementation of a two-time step, four-qubit quantum channel approximation using five ancilla qubits by iteratively applying the Stinespring unitary to the computational qubits and a new set of ancilla qubits every iteration. 1) In the quantum processing unit (QPU) the Stinespring unitary is performed to evolve $\rho_0$ to $\Phi_{\Delta t}(\rho_0)$. 2) Using movable tweezers, the entangled ancillas are deposited in the storage, and new ancillas are brought in from the reservoir. 3,4) The processes of 1) and 2) are repeated to evolve $\Phi_{\Delta t}(\rho_0)$ to $\Phi_{2\Delta t}(\rho_0)$.}
            \label{fig:neutralatom}
        \end{figure}    
    \end{minipage}
\end{widetext2}

\textcolor{white}{.}
\newpage
\textcolor{white}{.}
\newpage

As mentioned in Sec.~\ref{sec:Stinespring}, before preparing $\rho(2\Delta t)$ from $\rho(\Delta t)$, the old set of ancillas needs to be stored so that it is isolated from the rest of the system and a new set of ancillas is required for the evolution from $t=\Delta t$ to $t=2\Delta t$. A neutral atom quantum computer is well suited for this quantum channel extrapolation method compared to other architectures, for three main reasons:
\begin{enumerate}
    \item[(A)] With ground-ground qubits, after application of $U[\theta]$ the ancilla qubits are back in the $\{|0\rangle,|1\rangle\}$ manifold, which is well isolated from its surroundings and can be kept stable for long periods of time (up to minutes for $^{88}$Sr \cite{srlifetimes}). This gives rise to a viable means of creating storage and reservoir arrays.  
    \item[(B)] Using movable tweezers, qubits can be coherently moved on timescales comparable to the execution time of the elementary gate set, as experimentally achieved in Ref.~\cite{Bluvstein_2022}.
    \item[(C)] The system is hugely scalable in the number of qubits, since the number of tweezers scales linearly with the laser power. Per evolution step $n\Delta t\rightarrow (n+1)\Delta t$, only the computational qubits and one set of ancilla qubits need to be controlled in the quantum processing unit (QPU) when applying the Stinespring unitary, so there is no requirement for extra control when extrapolating to further time steps. {In other words, the repeated applications do not change the dimensionality of the parameter space. These extra ancilla qubits are only used for the time extrapolation steps and therefore do not play a role in scalability issues arising from barren plateaus.}
\end{enumerate}

An illustration of a practical implementation of the quantum channel extrapolation algorithm using a neutral atom quantum computer can be seen in Fig.~\ref{fig:neutralatom}.

\section{Results}
\label{sec:results}
{In this section, we present our VQA method for quantum channels using classical simulations for various target quantum channels. The code for these simulations can be found in an online repository \cite{QCA_Code} \footnote{The quantum channel learning problems presented in this work can be run on a high-end laptop or desktop.}. For each target quantum channel we explore, there are several steps necessary to arrive at the results presented here. First, the high-accuracy quantum state solver QuTiP \cite{qutip} simulates the target quantum channel to create the corresponding training data and the full quantum channel later used for error analysis. Secondly, we fix the ansatz and the corresponding parameters for our unitary dilation $U[\theta]$ and run either a pulse- or gate-based gradient descent as described in Sec.~\ref{sec:vqas}. Finally, once training has been performed, we analyze the difference between the quantum channel that arises from the learned unitary $U[\theta_{\text{opt}}]$ and the target quantum channel as calculated by QuTiP. For this comparison, we focus on extrapolation (as in~\eqref{eq:multitimestep}) to times $t$ significantly larger than that used in the training data, to better characterize how well the learned dilation captured the full behavior of the target quantum channel. Given that we only train on a few time steps, we expect the approximation to deviate more and more as the extrapolation progresses away from the largest training time step.}\\

{To illustrate the quality of the approximations, we show the extrapolation and the high-accuracy numerical results for some individual instances. Furthermore, we use the Bures distance as given by}

\begin{equation}
    d_{\text{Bures}}(\rho, \tilde{\rho})^2 = 2 \left(1-\operatorname{Tr}\left[\left(\sqrt{\rho} \tilde{\rho} \sqrt{\rho} \right)^{\frac{1}{2}}\right] \right)\leq 2.
\end{equation}
{
for a more quantitative comparison. For each quantum channel that we investigate, we calculate the average Bures distance between the approximation and the numerical solution for 10 pure initial states randomly drawn from the Haar measure \cite{haar}.}\\

{Various parameters have to be fixed to perform the numerical simulations and align the simulation settings with experimental setups. For the different target quantum channels, we set a characteristic timescale of $T_{\text{tqc}}$, which is independent of the quantum computer's state preparation timescale $\tau_f$. We choose to measure the quantum channel against every Pauli string, thus fully characterizing the quantum state at every discrete time step. The pulse-based system has a driving Hamiltonian $H_V$ determined by the Van der Waals interaction and control Hamiltonians $H_c$ that correspond to control over the coupling and detuning of the laser (cf.\ Appendix~\ref{app:rydberg}). The Van der Waals (VdW) interaction is set for $R=1\, \mu$m between nearest neighbor qubits and $C_6=0.422\,$kHz/$\mu$m$^6$ such that $V=0.422\,$ kHz between nearest neighbors. The continuous controls over the coupling and detuning of the control laser for all individual atoms are discretized as equidistant step functions. Together, this specifies $H_V$ and $H_c$ as defined in Sec.~\ref{sec:pulsebasedopt}.}\\

{ In Secs.~\ref{sec:1qubitdecay} and~\ref{sec:twoqubit}, we will approximate and analyze quantum channels on one and two qubit systems, respectively. Moreover, Sec.~\ref{sec:twoqubit} will feature examples of channels on quantum systems not native to the quantum computers on which they are analyzed. In Sec.~\ref{sec:pulsebasedgatebased}, we compare and contrast gate- and pulse-based methods. From this, we will learn that the quantum channel approximation method favors pulse-based approaches, which is why we preemptively showcase these results in the rest of this section.}

\subsection{1 Qubit decay}
\label{sec:1qubitdecay}
The evolution of a single qubit decaying from $\ket{1}$ to $\ket{0}$ with rate $\gamma_{\text{tqc}}$ and undergoing Rabi oscillations with frequency $\Omega_{\text{tqc}}$ is given by the Lindbladian, as in \eqref{eq:lindblad},
with
\begin{gather*}
H = \frac{\Omega_{\text{tqc}}}{2} \begin{pmatrix} 0 & 1 \\ 1 & 0 \end{pmatrix},\,\,
\Gamma_0 = \begin{pmatrix} 0 & 1 \\ 0 & 0 \end{pmatrix},\,\,
\rho(0) = \begin{pmatrix} \rho_{00} & \rho_{01} \\ \overline{\rho_{01}} & 1-\rho_{00} \end{pmatrix}.
\end{gather*}
The dynamics of this system allow for the state to be described analytically (cf. App.~\ref{app:1qubitdecay}), allowing for exact calculations of the errors of our quantum channel approximation. By the Stinespring dilation theorem (cf.\ Sec.~\ref{sec:Stinespring}), at most two ancilla qubits are necessary to approximate {quantum channels of this dimension. As this specific channel has one jump operator, it can be dilated using only one ancilla qubit.} We want to approximate single qubit decay with $\gamma_\text{tqc}=\Omega_\text{tqc}=0.5$, in units of $1/T_{\text{tqc}}$. From Fig.~\ref{fig:1qubitdecay}, we see that after minimizing the loss function (cf.~\eqref{eq:lossfunc}), the quantum channel approximation becomes extremely accurate. Note that the Hamiltonian splitting leads to significantly faster convergence of the unitary learning. The Bures error averaged over 10 new initial states on the first time step is $3.6 \cdot 10^{-4}$ and rises to $6.9 \cdot 10^{-4}$ after 10 re-applications of the unitary circuit.  From the figure and the evolution of the average Bures error, we see that the behavior of the quantum channel can be accurately extrapolated for longer times without the error increasing significantly. This is especially interesting because no knowledge on the system after $t_\text{train}$ is assumed.\\

\begin{widetext2}
    \begin{minipage}[b]{\linewidth}
    \begin{figure}[H]
        \centering        \includegraphics[width=\linewidth]{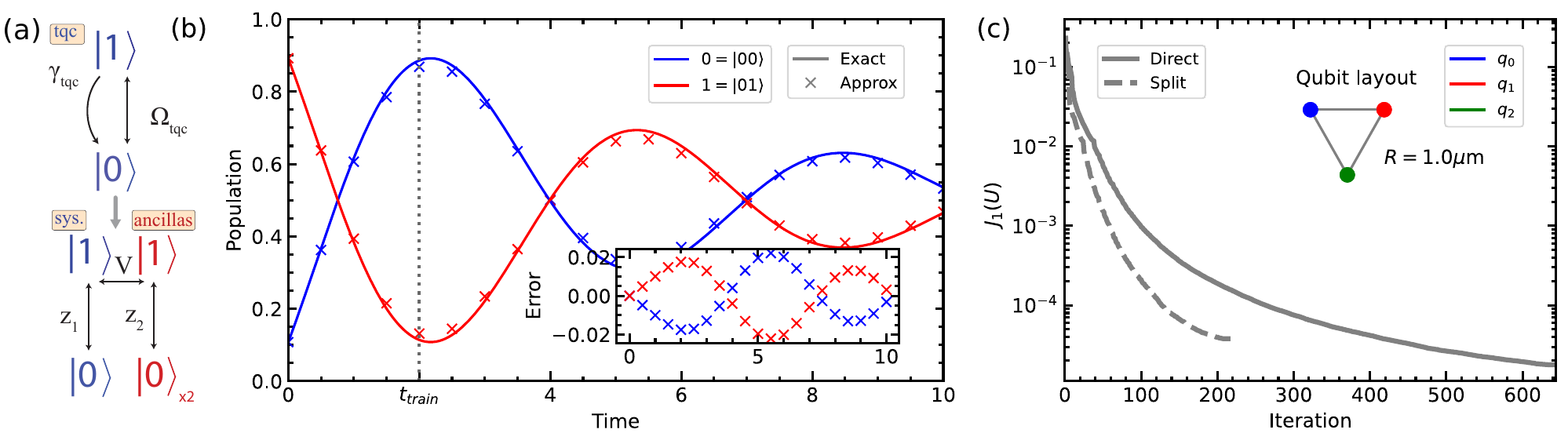}
        \caption{Quantum channel approximations for a target quantum channel describing single qubit decay with $\gamma_\text{tqc}=0.5$ and $\Omega_\text{tqc}=0.5$, both in units of $1/T_{\text{tqc}}$. The algorithm is trained on 10 deterministic initial states combined with all 4 Pauli strings to give $L=40$. Four training time steps are used, so that we have $t_\text{train}=4\Delta t$. Two ancilla qubits are taken, with all qubits positioned in an equilateral triangle. (a) Exact and approximated populations (with errors) for a single state not in the training data set with the pulses found by first learning the Hamiltonian part and then the decay. (b) Convergence of error $J(U)$ for separately learning the Hamiltonian part and the decay part (Split) and for directly training the full channel in one go (Direct). Includes the physical orientation of the qubits.}
        \label{fig:1qubitdecay}
    \end{figure}
    \end{minipage}
\end{widetext2}

\begin{widetext2}
    \begin{minipage}[b]{\linewidth}
    \begin{figure}[H]
            \centering
            \hypertarget{fig:1qubitpmdecay}{}
            \includegraphics[width=\columnwidth]{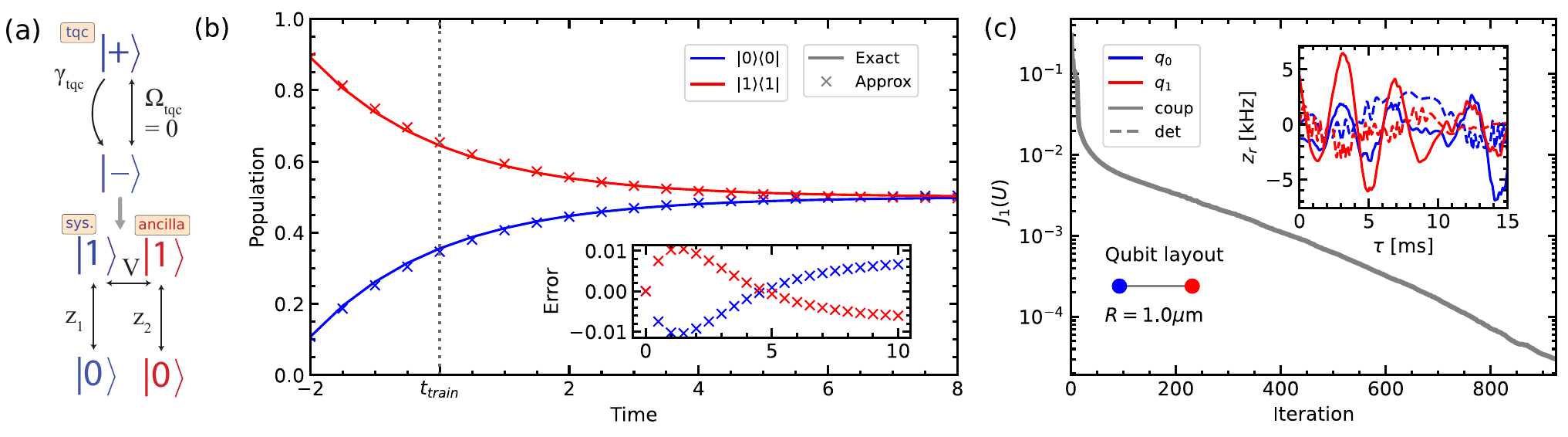}
            \caption{Quantum channel approximations for a target quantum channel describing single qubit decay from the $\ket{+}$ state to the $\ket{-}$ state with $\gamma_\text{tqc}=0.5$ and $\Omega_\text{tqc}=0$, both in units of $1/T_{\text{tqc}}$. The algorithm is trained on 10 randomly sampled initial states combined with all 4 Pauli strings to give $L=40$. Four training time steps are used, so that we have $t_\text{train}=4\Delta t$. One ancilla qubit is taken. (a) Quantum channel mapping. (b) Exact and approximated state populations for a single state not in the training data set. (c) Convergence of error $J(U)$ together with the final pulses on both qubits and the physical orientation of the qubits.}
            \label{fig:1qubitpm}
    \end{figure}    
    \end{minipage}
\end{widetext2}

\textcolor{white}{.}
\newpage
\textcolor{white}{.}
\newpage

A slight rephrasing of this problem will have the single qubit decay from the $\ket{+}$ to the $\ket{-}$ state, the eigenstates of the $\sigma_x$ Pauli string. This is a unitary transformation of the previous problem, which increases the difficulty as the quantum channel is no longer described in the native control basis of the $\ket{0}$ and $\ket{1}$ state. The results in Fig.~\ref{fig:1qubitpm} show this increase in difficulty by the increase in training time to achieve the same error $J$. The average Bures distance after the first time step is $9.4\cdot 10^{-5}$ and decreases to $6.1 \cdot 10^{-5}$ after 10 time steps. The increasing quality of the approximation is due to the approximation and the numerical solution converging to the same asymptote. Lastly, the pulses retrieved are relatively smooth and implementable in a real system.

\subsection{2-Qubit systems}
\label{sec:twoqubit}
{For more interesting systems, we also consider quantum channels that can be mapped to two qubits. In this section, we will only show systems where $H=0$, as the Hamiltonian part can be trained separately (see Sec.~\ref{sec:BarrenPlateauHamiltonianSplitting}) and the Hamiltonian behavior can obfuscate the quality of the approximation of the decay channels. We choose settings for which the qubits decay to a single basis state, as heuristically we find that the extremity of these steady states makes these channels hard to learn, thus giving the clearest picture of the capabilities and limitations of our method.}\\

As we only investigate systems with two or three decay terms, we correspondingly only need one or two ancilla qubits by the Stinespring dilation theorem (cf.\ Sec.~\ref{sec:Stinespring}). Nevertheless, to open up the search space, {it can be beneficial to increase the number of ancilla qubits. Although training can be performed with arbitrary initial states, we have manually chosen a set of training data to optimally use our available simulation power. The chosen set of training data consists of the 4 pure basis states $\ket{00}, \ket{01}, \ket{10}, \ket{11}$ and the 6 pure states that are given by the superposition of any two basis states. The evolution of these states efficiently captures the full behavior of the quantum channel with the least amount of training data.}\\

The first setting we consider is the 2-qubit case with individual qubit decay. The decay rates are taken as $\gamma_{0, \text{tqc}} = 0.5$ and $\gamma_{1, \text{tqc}} = 0.3$, all in units of $1/T_{\text{tqc}}$, and the interaction strength is $V_{\text{tqc}}=0 \,[1/T_{\text{tqc}}]$. The resulting behavior and approximation can be seen in Fig.~\hyperlink{fig:2qubitdecay}{\ref{fig:2qubitdecay}}.\\

From Fig.~\ref{fig:2qubitdecay}, we see that learning the target quantum channel becomes harder as the Pauli trace errors remain larger than in the single qubit results of Sec.~\ref{sec:1qubitdecay}. One can clearly see this in Fig.~\ref{fig:2qubitdecay}a, as the populations are no longer precisely extrapolated. Furthermore, the average Bures error on the first time step is $6.6 \cdot 10^{-3}$ and $1.8\cdot 10^{-2}$ after 10 time steps, which is more than an order of magnitude higher than for the single qubit target quantum channel. {If we were to fully separate the two computational qubits with their own ancilla qubit, this setting would essentially simulate the 1 qubit decay setting twice. However, the always-on interaction between the system qubits makes learning the correct channel more difficult, as the pulses have to simultaneously create the correct unitary dilation, while avoiding and counteracting the entanglement between the system qubits.} Despite this, it is remarkable to see that the qualitative behavior of the evolution remains well predicted, even far beyond the training time $t_{\text{train}}$.\\

\begin{widetext2}
\begin{minipage}[b]{\linewidth}
\begin{figure}[H]
    \begin{centering}
    \begin{minipage}[t]{\textwidth}
         \hypertarget{fig:2qubitdecay}{}
    \includegraphics[width=\columnwidth]{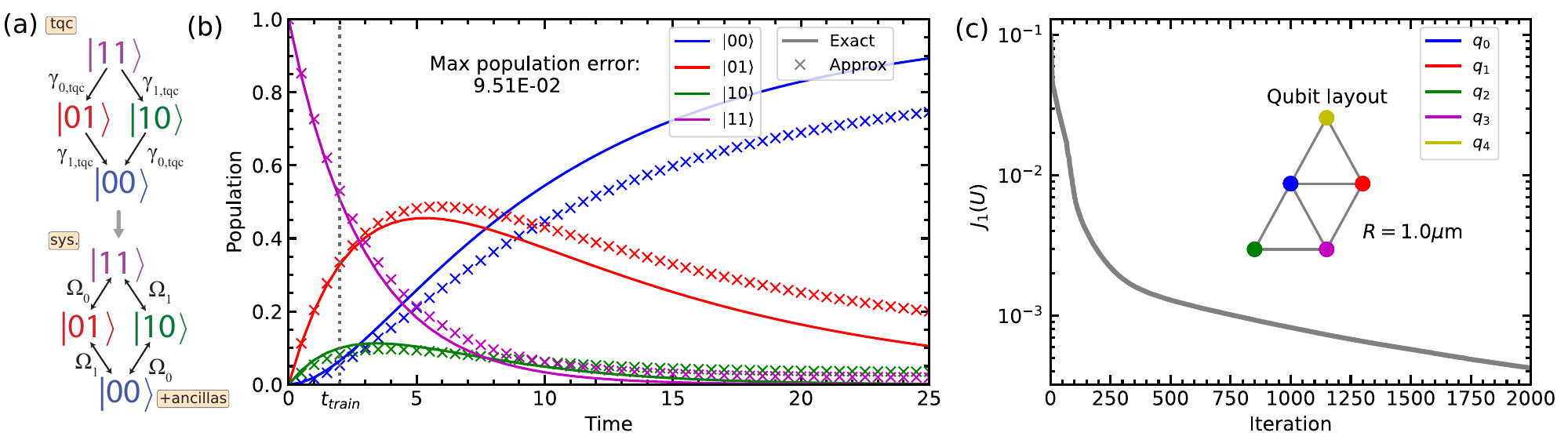}
    \end{minipage}

    \caption{Quantum channel approximations for a quantum channel describing two decaying qubits with interaction. The parameters of the target quantum channel are $\gamma_{0,tqc}=0.5$, $\gamma_{0,tqc}=0.3$ and an interaction strength of $V_{tqc} = 0$, all in units of $1/T_{\text{tqc}}$. The algorithm is trained on the 10 fixed initial states given combined with the 16 Pauli strings to give $L=320$. Three ancilla qubits are taken, positioned as illustrated in the inset in (b). (a) Quantum channel mapping. (b) Exact and approximated populations for the initial state $\rho = \ket{10}\bra{10}$, which is included in the training data set, to show the different decay rates. (c) Convergence of error $J(U)$ together with the physical orientation of the qubits.}
    \label{fig:2qubitdecay}
    \end{centering}
\end{figure}
\end{minipage}
\end{widetext2}

\textcolor{white}{.}
\newpage
\textcolor{white}{.}
\newpage

{The next two examples show quantum processes that are not native to the quantum computer they are approximated on and thus require a mapping between them. First, we analyze a 4-level system mapped to the 2-qubit states. The mapping is given by $\ket{3}\to \ket{10}, \ket{2} \to \ket{11}, \ket{1} \to \ket{01}, \ket{0} \to \ket{00}$. This ensures that every decay channel only flips one qubit. For the simulation, the 4-level system has the three decay channels from $\ket{3} \to \ket{2}$, from $\ket{2} \to \ket{1}$, and from $\ket{1} \to \ket{0}$, with rates $\gamma_{3 \to 2} = 0.5$, $\gamma_{2 \to 1}=0.4$, and $\gamma_{1 \to 0}=0.3$. The mapping to the 2 qubits results in three conditional decay channels, for example, qubit 2 decays from $\ket{0}$ to $\ket{1}$ if the first qubit is in the $\ket{0}$ state. These conditional decay channels are not easily implementable on the hardware. In App.~\ref{app:1qubitdecay}, we show a gate-based approach that can implement these conditional decay channels by using conditional SWAP-gates. The resulting training and approximation for this system can be seen in Fig.~\ref{fig:4level}.}\\

{From Fig.~\ref{fig:4level}, we see that learning the target quantum channel is more difficult than learning the normal 2-qubit decay system. The long-time extrapolations in both systems are of similar scale, but the 4-level system already makes noticeable errors within the limits of the training time steps. Furthermore, the average Bures error on the first time step is $5.3 \cdot 10^{-3}$ and $3.1\cdot 10^{-2}$ after 10 time steps. Although one can read some incorrect decay from $\ket{3}=\ket{01} \to \ket{0} = \ket{00}$, this rate is relatively low. The qualitative behavior is captured by the approximation, as can be seen in how the system decays sequentially through the different levels due to the slower decay rates at the lower levels. This is remarkable, as these quantum channels are not native to the quantum computer hardware.}

\begin{widetext2}
    \begin{minipage}[b]{\linewidth}
\begin{figure}[H]
\begin{centering}
    \begin{minipage}[t]{.95\textwidth}
         \hypertarget{fig:2qubitdecayb}{}
         \includegraphics[width=\columnwidth]{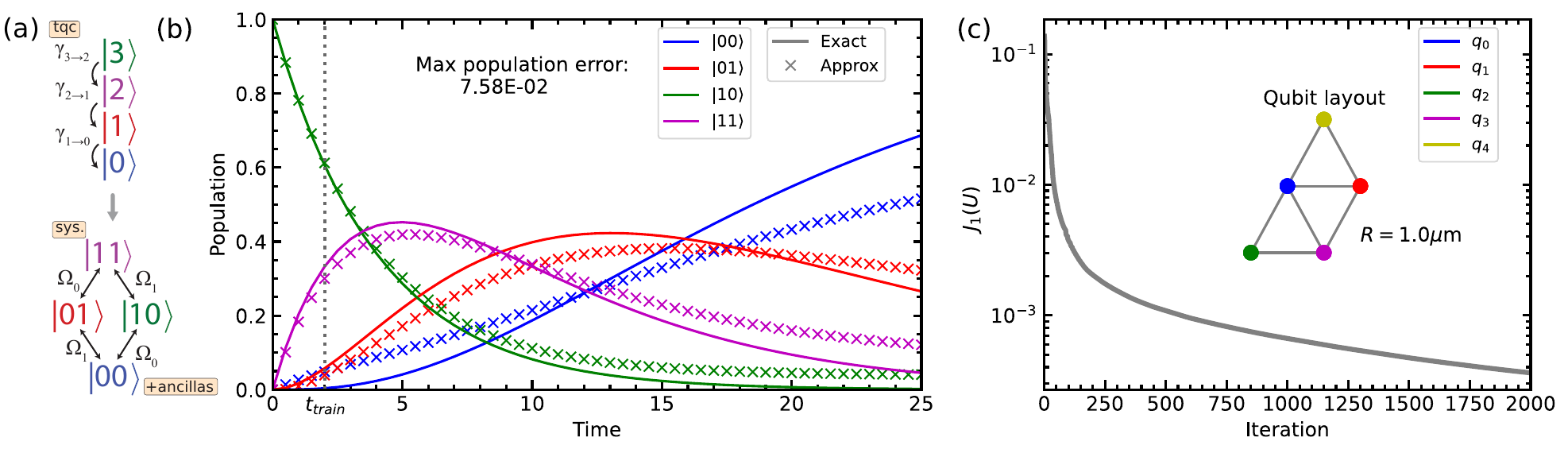}
    \end{minipage}%
\caption{Quantum channel approximations for a 4-level system mapped to a 2-qubit system with decay rates $\gamma_{3 \to 2} = 0.6$, $\gamma_{2 \to 1} = 0.5$, and $\gamma_{1 \to 0} = 0.4$ in units of $1/T_{\text{tqc}}$. For the approximating qubit system, we take the same parameters and training setup as in Fig.~\ref{fig:2qubitdecay}. (a) Exact and approximated populations for the initial state $\rho = \ket{10}\bra{10}$, which is included in the training data set, to show the sequential decay. (b) Convergence of error $J(U)$ together with the physical orientation of the qubits.}
    \label{fig:4level}
    \end{centering}
\end{figure}
    \end{minipage}
\end{widetext2}

\begin{widetext2}
    \begin{minipage}[b]{\linewidth}
    \begin{figure}[H]
        \centering
        \includegraphics[width=\columnwidth]{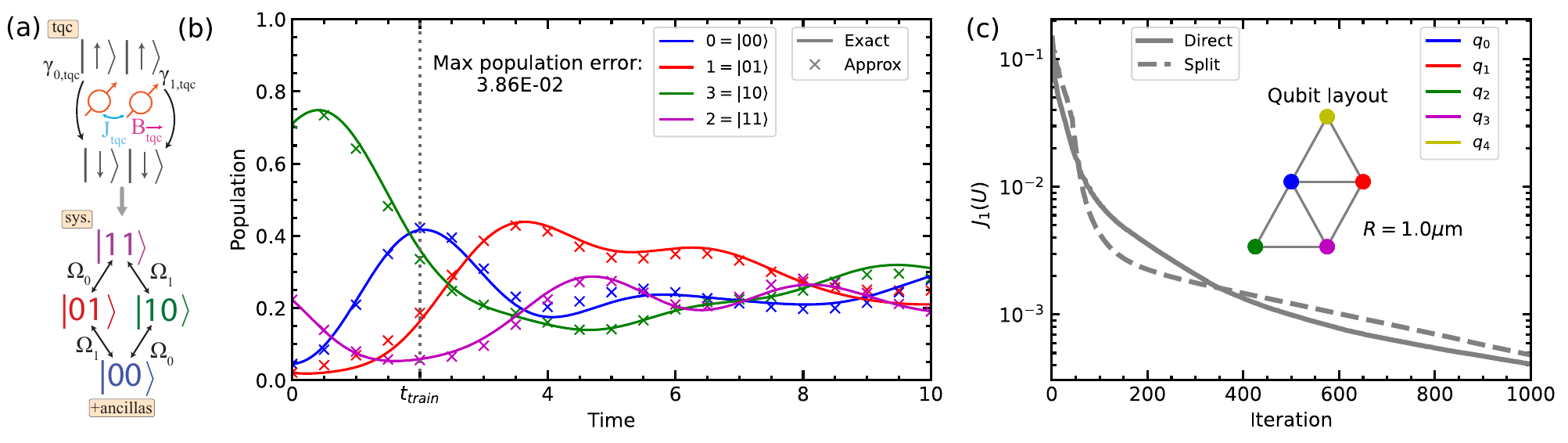}
        \caption{Quantum channel approximations for a 2-qubit TFIM model with $B_\text{tqc}=0.5$, $J_\text{tqc}=0.4$, $\gamma_{0,\text{tqc}}=0.5$, and $\gamma_{1,\text{tqc}} = 0.3$, in units of $1/T_{\text{tqc}}$. For the approximating qubit system, we take the same parameters and training setup as in Fig.~\ref{fig:2qubitdecay}. (a) Exact and approximated populations for a single state not included in the training data set for split learning (b) Convergence of error $J(U)$ for learning the Hamiltonian behavior and decay channels separately (split) or simultaneously (direct), together with the physical orientation of the qubits.}
        \label{fig:ising}
    \end{figure}
    \end{minipage}
\end{widetext2}

\textcolor{white}{.}
\newpage
\textcolor{white}{.}
\newpage

The third and last 2-qubit system that we analyze is the transverse field Ising model (TFIM) with decay \cite{tfim}. In this model, spins interact with a transverse magnetic field as well as with their nearest neighbors. For spins aligned in a straight line, the TFIM Hamiltonian takes the form
\begin{equation}
    H =- B_\text{tqc} \sum_i X_i+J_\text{tqc} \sum_i Z_i Z_{i+1},
\end{equation}
with $Z_i$ and $X_i$ respectively the Pauli $Z$-operator and Pauli $X$-operator on qubit $i$, $J_\text{tqc}$ the coupling strength, $\gamma_{i,\text{tqc}}$ the decay on qubit $i$, and $B_\text{tqc}$ the transverse external field strength, both in units of $1/T_{\text{tqc}}$. Fig.~\ref{fig:ising} shows results for a 2 spin TFIM. {The average Bures error on the first time step is $1.0 \cdot 10^{-2}$ and $7.4\cdot 10^{-3}$ at 10 time steps, again with faster initial convergence for Hamiltonian splitting, indicating advantageous properties for barren plateau mitigation.} Similar to the 2-qubit case of Fig.~\ref{fig:2qubitdecay}, we see increasing errors for higher extrapolation times, but good prediction of the general behavioral trend of the evolution. {The system with decay from $\ket{+}$ to $\ket{-}$, the 4-level system and the TFIM model show that the algorithm can approximate target quantum channels which are not native to the quantum computing system they are approximated on.}

\subsection{Pulse-based vs. gate-based}
\label{sec:pulsebasedgatebased}
In the current NISQ-era, gate-based optimization methods are most widely implemented. In recent work, it has been suggested that stochastic and pulse-based methods can lead to higher expressibility and faster convergence of optimization \cite{pulsesandbackagain,deKeijzer2023pulsebased,stochastic1,stochastic2}. For our algorithm, we compare gate-, stochastic gate-, and pulse-based methods using the notion of equivalent evolution processes \cite{deKeijzer2023pulsebased}, which considers that for all methods, the system evolves under the most similar circumstances. Concretely, this means all methods are run on a quantum computation system with similar controls, entanglement operations, and evolution times.\\

We consider control over the coupling strength $\Omega$ for the pulse-based method and a hardware efficient ansatz with $ZXZ$ parametrized gates for the gate-based method, both resulting in full rotational control of the Bloch sphere \cite{deKeijzer2023pulsebased}. Furthermore, we consider a Van der Waals drift Hamiltonian $H_V$ and entanglement gates $U_{ent}=\exp(-iH_V\tau_V)$ for the gate-based method. {The gate-based ansatz consists of 10 blocks of $ZXZ$ gates, with 9 entanglement gates $U_{ent}$ in between.} We assume $V=0.07$ kHz with $R=1 \mu$m and $C_6=0.07$ kHz$/\mu$m$^6$. In order to supply enough time for entanglement, we take $\tau_V=1/V=10$ ms and take Rabi frequencies in the order of 1 kHz to get $\tau_g=1$ ms. To obtain similar evolutions, we take $\tau_{f,\text{pulse}}=\tau_{f,\text{gate}}$, which completes the construction of the equivalent evolution processes. To add stochasticity to the gate-based optimization, a random subset of gates is chosen to update the parameters for each iteration, instead of optimizing the entire gate set.\\

\begin{figure}[t]
    \centering
    \includegraphics[scale=0.56]{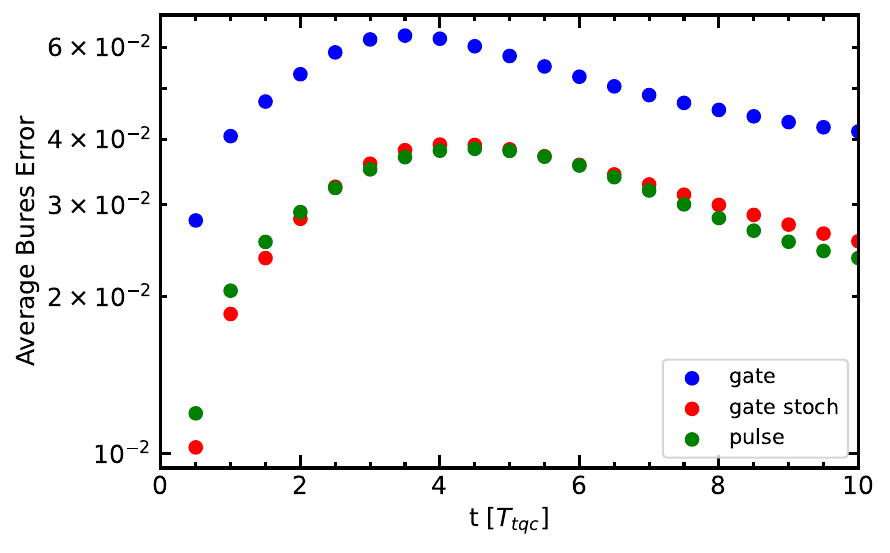}    
    \caption{Comparison of the evolution over time of the Bures error for a gate-based, a stochastic gate-based, and a pulse-based approach. The target quantum channel is a decay model on two qubits with $\gamma_{0,tqc}=0.3$ and $\Omega_{0,tqc}=0.5$ for qubit 0, $\gamma_{1,tqc}=0.2$ and $\Omega_{1,tqc}=0.35$ for qubit 1 and an interaction strength of $V_{tqc} = 0.2$, all in units of $1/T_{\text{tqc}}$. The Bures error is the average of the extrapolations on 10 new states. Training done using the notion of equivalent evolutions and for the same number of quantum evaluations $\#\text{QE}$.}
    \label{fig:gatevspulse}
\end{figure}

In Fig.~\ref{fig:gatevspulse}, we compare the average Bures distances over time for the unitary approximations based on a gate-based, a stochastic gate-based and a pulse-based method, each trained for the same number of $\#\text{QE}$. The target quantum channel is again a decay model on two qubits with $\gamma_{0,tqc}=0.3$ and $\Omega_{0,tqc}=0.5$ for qubit 0, $\gamma_{1,tqc}=0.2$ and $\Omega_{1,tqc}=0.35$ for qubit 1 and an interaction strength of $V_{tqc} = 0.2$, all in units of $1/T_{\text{tqc}}$. The pulse-based method and the stochastic gate-based method perform very similarly, while both outperform the gate-based method. We hypothesize that the pulse-based method can be improved by adding stochasticity as well. Note that this comparison was made on a data set where all three methods converged to a reasonable endpoint. Generally, we find that the gate-based methods had problems with finding good local minima of the loss function for a large fraction of randomized sets of training data.

\section{Conclusion}
\label{sec:conclusion}

In this work, we introduce an algorithm for quantum channel approximation and extrapolation. This pulse-based method differentiates itself from previous work by being able to approximate the quantum channel based purely on measurement data, and the inclusion of multiple time steps. The method variationally learns the Stinespring unitary describing the quantum channel at a fixed time $\Delta t$, either through a gate- or pulse-based method. The learning of the quantum channel seems to benefit from a Hamiltonian splitting ansatz, indicating beneficial properties for barren plateau mitigation. This approximation can later be extrapolated to make future time predictions for the quantum channel behavior at discrete multiples of $\Delta t$. The method has, through simulations, shown proof of concept for non-trivial target quantum channels based on NISQ qubits, multi-level systems, and Ising models. An analysis between gate- and pulse-based implementations of the algorithm, using equivalent evolution processes, has shown that adding stochasticity or switching to a pulse-based method is beneficial for approximating quantum channels given small state preparation times, which is especially important for mitigating decoherence in the NISQ era. \\

Furthermore, it is reasoned that a neutral atom quantum computing system is very well suited for executing this algorithm. The algorithm requires numerous ancilla qubits that need to be coherently moved and stored without being controlled, to which neutral-atom systems are adapted given their high scalability, long coherence times, and modifiable qubit topologies.\\

In future work, a direction of improvement would be in using prior information on the quantum channel to improve convergence. We hypothesize that by knowing the Kraus decomposition \cite{lindblad1} or steady state, it could be possible to construct a set of observables and initial states that provide more information on the channel's behavior. Furthermore, we would like to include a more sophisticated learning algorithm than a simple gradient descent. One promising direction lies in incorporating a Wasserstein distance into the loss function, as this has been shown to significantly reduce the barren plateau problem in state preparation \cite{Kiani_2022}. Another interesting direction would be to implement pulses from the Fourier domain, as this has the potential to mitigate the barren plateau as well \cite{Broers24}.

\textcolor{white}{.}
\newpage

\section*{Data Availability}
The data that support the findings of this study are available from the corresponding author upon reasonable request.

\section*{Code Availability}
The code that support the findings of this study are available from an online repository \cite{QCA_Code}.

\section*{Acknowledgements}
We thank Jasper Postema and Jurgen Snijders for fruitful discussions. We thank David Chen for his work on the simulation code. This research is financially supported by the Dutch Ministry of Economic Affairs and Climate Policy (EZK), as part of the Quantum Delta NL program, and by the Netherlands Organisation for Scientific Research (NWO) under Grant No.\ 680.92.18.05. L.Y. Visser and O. Tse acknowledge support from NWO grant NGF.1582.22.009.

\vfill\null

\bibliographystyle{apsrev4-1}
\bibliography{Bibliography.bib}

\newpage
\onecolumngrid

\appendix
\section{Multiple time step methods}
\label{app:multistep}

A possibility of improving the learning capabilities of our method is training with data in multiple time steps. If measurements on $\Phi_{n\Delta t}(\rho_0)$ for $n \in [1,2,\cdots, N]$ are known, the loss function can be redefined as
\begin{equation}\label{eq:j1 error incl time}
    J(U) = \sum_{n=1}^N \sum_{l=1}^L \Tr_A[O_l (\tilde{\rho}_{l,n} - \rho_{l,n})]^2,
\end{equation}
with 
\begin{equation}
    \rho_{l,n} = \Phi_{n\Delta t}(\rho_{l,0}) = \underbrace{\Phi_{\Delta t} ( \Phi_{\Delta t} ( \cdots \Phi_{\Delta t}(\rho_{l,0}) \cdots ))}_{n  \text{ repeated applications}} \text{, and }
    \tilde{\rho}_{l,n} = \Tr_\mathcal{B}[U \cdot \tilde{\rho}_{l,n-1} \otimes \ket{0}\bra{0}_\mathcal{B} \cdot U^\dag], \qquad \tilde{\rho}_{l,0} = \rho_{l,0}.
\end{equation}
{Measuring at multiple time steps foremost creates more pairs of input-output data on the quantum channel. If measuring the observables fully determines the density matrix, this is the only dominant contribution of adding multiple time steps. However, repeated applications can be used to infer information about the channel that is not captured directly by the observables. In a single application of the quantum channel $\widehat{\Phi}$, the part of the density matrix that is orthogonal to the observables $O_l$ can be freely distributed over the orthogonal space because the loss function does not distinguish there. However, in repeated applications of the channel, this orthogonal part can be mapped to the measurable part. Thus, by repeated application of the quantum channel, the loss function has the opportunity to capture the dynamics of the part that is not measured by the observables.}

Straightforward variational differentiation in the pulse-based case shows that the Fr\'echet derivative $\frac{\delta}{\delta U}J$ for two time steps ($n=2$) is given by
\begin{equation}
\begin{aligned}
    \frac{\delta J}{\delta U} &= 4 \delta(\tau_f-\tau) \sum_{l=1}^L \Big(
    \Tr_A[O_l (\tilde{\rho}_{l,1} - \rho_{l,1})]\times\left( O_{l} \otimes I_\mathcal{B} \cdot U \cdot \tilde{\rho}_{l,0}\otimes \ket{0}\bra{0}_\mathcal{B}\right)  \\
    & + \Tr_A[O_l (\tilde{\rho}_{l,2} - \rho_{l,2})]\times \left( O_{l} \otimes I_\mathcal{B} \cdot U \cdot \tilde{\rho}_{l,1}\otimes \ket{0}\bra{0}_\mathcal{B}\right)  \\
    & + \Tr_A[O_l (\tilde{\rho}_{l,2} - \rho_{l,2})]\times \Tr_\mathcal{B}[U^\dag \cdot O_{l} \otimes I_\mathcal{B} \cdot U \cdot I_A \otimes \ket{0}\bra{0}_\mathcal{B}] \otimes I_\mathcal{B} \cdot U \cdot \tilde{\rho}_{l,0}\otimes \ket{0}\bra{0}_\mathcal{B}
    \Big).
\end{aligned}
\end{equation}

This can be extended to show that the Fr\'echet derivative $\frac{\delta}{\delta U}J$ for $n$ time steps is given by
\begin{equation}
    \frac{\delta J}{\delta U} = 4 \delta(\tau_f-\tau) \sum_{n=1}^N \sum_{l=1}^L \sum_{k=0}^{n-1} \Tr_A[O_l (\tilde{\rho}_{l,n} - \rho_{l,n})] \times \left(O_{l,k} \otimes I_\mathcal{B} \cdot U \cdot \tilde{\rho}_{l,n-k-1}\otimes \ket{0}\bra{0}_\mathcal{B}\right),
\end{equation}
with $O_{l,k}$ defined as
\begin{equation}
    O_{l,k} = \Tr_\mathcal{B}[U^\dag \cdot O_{l,k-1} \otimes I_\mathcal{B} \cdot U \cdot I_A \otimes \ket{0}\bra{0}_\mathcal{B}], \qquad O_{l,0} = O_l.
\end{equation}

This gradient is very similar to the one in~\eqref{eq:frechet J1}. However, finding a way to measure $O_{l,k}$ as observable is not directly possible. Instead, $O_{l,0}$ has to be measured w.r.t. the entire entangled system, including all previously entangled sets of ancillas. The number of quantum state evaluations to calculate the gradient for multiple time steps is increased by a factor $\frac{1}{2}N(N-1)$ over the single time step method, for a total of $\#\text{QE}=\frac{1}{2}N(N-1) K L R$ quantum state evaluations per gradient calculation.

\section{Rydberg Physics}\label{app:rydberg}
This section introduces basic Rydberg physics to identify what drift and control Hamiltonians  as in~\eqref{eq:hamiltoniaoc} can look like on a Rydberg atom quantum computing system. This also yields candidates for the control operators $Q_r$, as in~\eqref{eq:controlhamiltonian4}. For this section, we assume ground-Rydberg qubits \cite{review1} such that the $|1\rangle$ state is a Rydberg state.\\

The Rydberg states have a passive `always-on' interaction, which is described by a drift Hamiltonian $H_V$, depending on the choice of qubits scheme \cite{rydbergqubit}, as a Van der Waals (VdW) interaction \cite{vdwaals} or a dipole-dipole interaction (dip) \cite{dipole}
\begin{equation}
\begin{aligned}
    H_{d,\text{VdW}}&=\sum_{i=1}^m\sum_{j=1}^m\frac{C_6}{R_{ij}^6}|11\rangle_{ij}\langle 11|_{ij},\\
    H_{d,\text{dip}}&=\sum_{i=1}^m\sum_{j=1}^m\frac{C_3}{R_{ij}^3}\left(|01\rangle_{ij}\langle 10|_{ij}+\text{h.c.}\right),
    \label{eq:rydbergvdwinteraction}
    \end{aligned}
\end{equation}
where $R_{ij}$ is the interatomic distance.  We also define the characteristic interaction as $V:=C_n/\min_{ij}{(R_{ij}^n)},\, n\in\{3,6\}.$\\

To perform single qubit manipulations on qubit $j$, a laser interacts with the atom to realize the Hamiltonian \cite{rydbergqubit,rydberg1,rydberg2}
\begin{equation}
\begin{aligned}
    H_{j}^{01}(\tau)=\frac{\Omega_{j}(\tau)}{2} \left(e^{i \varphi_{j}(\tau)}|0\rangle_j\langle 1|_{j}+e^{-i \varphi_{j}(\tau)}|1\rangle_j\langle 0|_{j}\right)-\Delta_{j}(\tau)|1\rangle_{j}\langle 1|_j.
\end{aligned}
    \label{eq:qubitlightinteraction}
\end{equation}
Here, $\Omega_j(\tau)$ denotes the coupling strength, $\varphi_j(\tau)$ the phase of the laser coupled to atom $j$, and $\Delta_j(\tau)$ = $\omega_j(\tau)-\omega_0$ the detuning of the laser frequency $\omega_j(\tau)$ from the energy level difference $\omega_0$. \\

The control Hamiltonian $H_{c}$ can contain several terms depending on which of the mentioned laser parameters can be controlled. In general, the control Hamiltonian takes the form 
\begin{equation}
    H_{c}[z(\tau)]=\sum_{r=1}^RQ_rz_r(\tau)+Q_r^\dagger \overline{z_r(\tau)},
    \label{eq:controlhamiltonian}
\end{equation}
where $z_r(\tau)\in \mathbb{C}$, and $Q_r$ is a $m$-qubit operator. The choice of $z_r(\tau)$ being a complex number stems from the fact that it could represent both the coupling strength and the phase in~\eqref{eq:qubitlightinteraction}. In this work, we assume no control over the phase such that in all cases $z_r(\tau)\in\mathbb{R}$. For the control Hamiltonian $H_c$
\begin{equation*}
    H_{c}[z]=H_{c}^{\text{coup}}[z^{\text{coup}}]+H_{c}^{\text{det}}[z^{\text{det}}],
\end{equation*}
with $z = (z^{\text{coup}}, z^{\text{det}}),$ $R=2m$. Here, the coupling control $H_{c}^{\text{coup}}$ and detuning control $H_{c}^{\text{det}}$ take the form of~\eqref{eq:controlhamiltonian} with
\[
    Q_r^{\text{coup}}=|0\rangle\langle1|_r\quad\text{and}\quad Q_r^{\text{det}}=|1\rangle\langle1|_r,
\]
respectively. Notice that having both coupling and detuning allows for full control on the Bloch sphere of each individual qubit. This is why this is also referred to as \textit{rotational control} \cite{rotational}. 

\section{Single qubit decay}
\label{app:1qubitdecay}
{This appendix illustrates an analytic example of a single qubit decay quantum channel together with its unitary dilation. This can be used for error analysis as well as insights in the Stinespring dilation process.}\\

A single qubit decaying from $\ket{1}$ to $\ket{0}$ with rate $\gamma$ and undergoing Rabi oscillations with frequency $\Omega$ is described by the Lindblad equation, with
\begin{gather*}
H = \frac{\Omega}{2} \begin{pmatrix} 0 & 1 \\ 1 & 0 \end{pmatrix},\qquad
\Gamma = \begin{pmatrix} 0 & 1 \\ 0 & 0 \end{pmatrix},\qquad \rho(0) = \begin{pmatrix} \rho_{00} & \rho_{01} \\ \overline{\rho_{01}} & 1-\rho_{00} \end{pmatrix}.
\end{gather*}
The dynamics of this system allow for the state to be described analytically as

\begin{equation}
\begin{aligned}
    \rho_{00}(t) &= B + (\rho_{00}-B) \exp\left(-\frac{3}{4}\gamma t \right) \left(\cosh\left(\frac{1}{4} t F\right)+ \frac{C}{F}\sinh\left(\frac{1}{4} t F \right) \right), \\
    \rho_{01}(t) &= B' + \operatorname{Re}\left(\rho_{01}\right) \exp\left(-\frac{1}{2} \gamma t\right) + \left(i \operatorname{Im}\left( \rho_{01}\right) - B' \right) \exp\left(-\frac{3}{4}\gamma t \right) \left(\cosh\left(\frac{1}{4} t F\right)+ \frac{C'}{F}\sinh\left(\frac{1}{4} t F \right) \right),
\end{aligned}
\end{equation}
with $ F =\sqrt{\gamma^2 - 16 \Omega^2}$ and
\begin{equation}
\begin{aligned}
    B = \frac{\gamma^2 + \Omega^2}{\gamma^2+2 \Omega^2}&, \qquad  C = -\gamma - \gamma \frac{4 \Omega^2}{(1-\rho_{00})\gamma^2 + (1-2\rho_{00})\Omega^2},\\ 
    B' = \frac{i \gamma \Omega}{\gamma^2 + 2 \Omega^2}&, \qquad C' = \gamma - 4 \Omega \frac{(1-\rho_{00})\gamma^2 + (1 - 2\rho_{00})\Omega^2}{\operatorname{Im}\left(\rho_{01}\right)(\gamma^2 + 2\Omega^2) -  \gamma \Omega }.
\end{aligned}
\end{equation}

For the single qubit system with $H= 0$, the solution is given by
\begin{equation}
    \rho(t) = \begin{pmatrix}
        \rho_{00}(0) + \rho_{11}(0)(1-\exp(-\gamma t))  & \rho_{01}(0) \exp(-\frac{1}{2}\gamma t) \\ 
    \rho_{10}(0) \exp(-\frac{1}{2}\gamma t)         & \rho_{11}(0)\exp(-\gamma t) 
    \end{pmatrix}.
\end{equation}

The unitary given by
\begin{equation}
    U_t =
    \begin{pmatrix}
        1 & 0 & 0 & 0 \\
        0 & \exp(-\frac{1}{2} \gamma t) & -\left(1-\exp(-\gamma t) \right)^{\frac{1}{2}} & 0 \\
        0 & \left(1-\exp(-\gamma t) \right)^{\frac{1}{2}} & \exp(- \frac{1}{2} \gamma t) & 0 \\
        0 & 0 & 0 & 1
    \end{pmatrix}
\end{equation}
satisfies
\begin{equation}
    \Phi_t(\rho) = \Tr_\mathcal{B}\left[U_t \cdot \rho \otimes \ket{0}_\mathcal{B} \bra{0}_\mathcal{B} \cdot U_t^\dagger\right],
\end{equation}
and can be implemented using a parameterized SWAP-gate.\\ 

Finally, we can find unitary dilations for other decay channels by unitary transformations and conditional extensions. Decay channels of the form $\ket{+} \to \ket{-}$ can be achieved via basis transformations. Decay channels of the form $\ket{11} \to \ket{10}$ can be achieved by making this unitary conditional on the first qubit being $|1\rangle$. Implementing multiple quantum channels using these dilations requires 1 qubit for every decay channel, which is significantly less efficient than the 1 decay channel per dimension of the ancilla space as given by the Stinespring Dilation theorem. Furthermore, these conditional unitary transformations are generally tough to implement for high-dimensional systems.

\end{document}